\documentclass[%
 reprint,superscriptaddress,
 amsmath,
 amssymb,
 aps
]{revtex4-2}

\usepackage{braket}
\usepackage{graphicx}
\usepackage{dcolumn}
\usepackage{bm}
\usepackage[export]{adjustbox}
\usepackage{color}
\definecolor{mblue}{RGB}{30,97,165}
\definecolor{mred}{RGB}{165,20,41}
\usepackage[colorlinks=true, linkcolor=mblue, citecolor=mblue, urlcolor=mblue, breaklinks=true]{hyperref}

\begin{document}

\title{Impurities in cryogenic solids: a new platform for hybrid quantum systems}

\author{Andrew N. Kanagin}
\email{andrew.kanagin@tuwien.ac.at}
\affiliation{Vienna~Center~for~Quantum~Science~and~Technology,~Atominstitut,~TU~Wien,~1020~Vienna,~Austria}

\author{Nikolaus de Zordo}
\affiliation{Vienna~Center~for~Quantum~Science~and~Technology,~Atominstitut,~TU~Wien,~1020~Vienna,~Austria}

\author{Andreas~Angerer}
\affiliation{Vienna~Center~for~Quantum~Science~and~Technology,~Atominstitut,~TU~Wien,~1020~Vienna,~Austria}

\author{Wenzel Kersten}
\affiliation{Vienna~Center~for~Quantum~Science~and~Technology,~Atominstitut,~TU~Wien,~1020~Vienna,~Austria}

\author{Nikolaos Lagos}
\affiliation{Vienna~Center~for~Quantum~Science~and~Technology,~Atominstitut,~TU~Wien,~1020~Vienna,~Austria}

\author{Jörg Schmiedmayer}
\affiliation{Vienna~Center~for~Quantum~Science~and~Technology,~Atominstitut,~TU~Wien,~1020~Vienna,~Austria}

\author{Elena S. Redchenko}
\email{elena.redchenko@tuwien.ac.at}
\affiliation{Vienna~Center~for~Quantum~Science~and~Technology,~Atominstitut,~TU~Wien,~1020~Vienna,~Austria}

\date{\today}

\begin{abstract}
Hybrid quantum systems offer a promising platform for studying quantum phenomena and developing applied technologies, benefiting from the individual strengths of their components. Here, we present a novel hybrid quantum platform composed of solid noble gas crystals doped with spin impurities atop superconducting resonators. The noble gas crystals provide a soft, inert, predominantly spin-0 host matrix for the atomic impurities, while the alkali atoms have addressable and long-lived hyperfine transitions in the GHz regime. We demonstrate the ability to reach the strong coupling regime between the atomic impurity ensemble and the superconducting resonator at mK temperatures, and perform coherence time measurements. Our proof-of-principle measurements show that this platform offers a unique architecture for exploring fundamental quantum effects and new quantum technologies.
\end{abstract}

\maketitle

Hybrid quantum systems leverage the combined strengths of individual systems to overcome problems that a single platform commonly encounters \cite{kurizki2015quantum}. One particularly promising class of hybrid platforms combines superconducting microwave circuits with solid-state spins \cite{clerk2020hybrid}. There, superconducting circuits provide fast control and readout \cite{oakes2023fast}, while spins naturally exhibit longer coherence times \cite{boyd2006optical, wang2025nuclear}. This approach has already enabled key milestones, including the observation of collective effects \cite{angerer2018superradiant,lei2023many}, the realization of microwave multimode storage \cite{Ranjan2020Multimodestorage}, and the detection of a single electron-spin-resonance \cite{wang2023single}. However, well-studied host materials -- from diamond to rare-earth oxides -- introduce decoherence via magnetic background, structural disorder, or surface noise \cite{le2021twentythreemllisecond, Bayliss2022EnchanceSpinCohernceHostMatrix}. Overcoming these limitations is essential for advancing the field toward scalable quantum technologies.

A less common but favorable host matrix for spins is one made from inert gases, such as helium, para-hydrogen, neon, etc \cite{bondybey1996new}. These gases have filled valence shells and, under normal conditions, do not react. Unlike the typical bonds found in nature (ionic, covalent, metallic), solid inert gases are held together by Van der Waals interactions, which are typically orders of magnitude weaker. As a consequence, solidification requires cryogenic temperatures ($\sim$10s of K) or high pressures. Solid inert gases serve as ideal substrates for single-electron qubits, as excess electrons naturally bind to their surface in vacuum \cite{zhou2022single,guo2025quantum,belianchikov2025image,tian2025nbtin}. They can also be doped with various impurity species, such as ions \cite{bondybey2001mass, redeker2017matrix,tsuge2018spectroscopy}, atoms \cite{hofmann1979matrix, braggio2022spectroscopy, battard2023cesium}, and reactive or polyatomic molecules \cite{becker1956spectroscopic, lee2006internal, park2017brute}, providing a test bed for studying different types of interactions. Importantly, impurities embedded in cryogenic solids remain relatively unaffected by the host matrix \cite{Pipkin_ArRb, cradock1975matrix, upadhyay2019spin, upadhyay2020ultralong}. Moreover, cryogenic crystals do not have adverse effects on superconducting materials \cite{Valenti2024}, making them suitable for hybrid integration. Finally, cryogenic solids can offer a reduction in the local magnetic noise, as a variety of rare gases have only stable isotopes with no nuclear spin.

Here, we present a novel hybrid quantum platform that integrates a doped cryogenic crystal with a superconducting resonator. This system, modular by design, enables diverse combinations of host matrices and dopant species. As a concrete realization, we employ neon (Ne) as the host, selected for its favorable melting point, and sodium (Na) as the dopant, chosen for its favorable hyperfine levels. At millikelvin temperatures, the superconducting resonator couples magnetically to the ground-state hyperfine transition of the embedded Na atoms, demonstrating a clear vacuum Rabi splitting between the atomic ensemble and the microwave cavity. We further measure the system’s relaxation and coherence times, establishing quantitative benchmarks for performance. These demonstrations position cryogenic solids with embedded impurities atop superconducting circuits as a versatile platform for next-generation hybrid quantum technologies.

\section{Crystal creation}

\begin{figure*}[t!]
    \includegraphics[width=1.95\columnwidth]{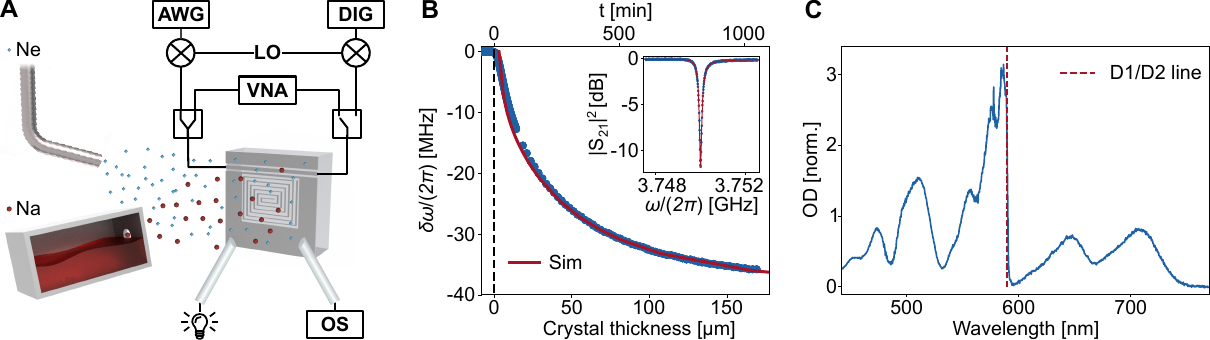}
    \caption{(\textbf{A}) Schematic illustration of the simplified experimental setup. Neon (Ne) gas diffusely exists the nozzle from inside the cryostat and sodium (Na) from the atomic beam oven mounted to the outer vacuum shield, condensing simultaneously on the surface of the superconducting resonator. Absorption spectroscopy of the crystal is done using two split fibers connected to the broadband light on the input and an optical spectrometer (OS) at the output. Continuous tone microwave measurements are performed using a vector network analyzer (VNA). The system gets continuous input from an RF source and pulsed input from the upconverted arbitrary waveform generator (AWG) signal. Analog downconversion and digitization (DIG) are used for coherence measurements. (\textbf{B}) Resonator frequency shift, $\delta\omega/(2\pi)=\omega_{\text{c}}(t)/(2\pi)-\omega_{\text{c}}(t=0)/(2\pi)$, as a function of crystal growth time. The start of the crystal growth is indicated with the black dashed line ($t = 0$). The resonator shift is fitted numerically with the Sonnet simulation result as a function of crystal thickness. The inset shows the fitted transmission spectrum of the cavity before the crystal is grown. (\textbf{C}) Optical transmission of a fully grown crystal in terms of the optical depth (OD) as a function of wavelength measured at 100\,mK. The OD is normalized to the off-resonance background OD after the crystal is grown. The largest OD is centered around 586\,nm and corresponds to the D1 and D2 doublet (red dashed line) of Na. The additional peaks are attributed to dimers and trimers of Na in the crystal.  
    }  
    \label{fig:ADR_crystal_growth}
\end{figure*}

We create the cryogenic crystal \textit{in situ} by depositing the host matrix gas and impurity atoms simultaneously atop the cooled superconducting circuit, which is held at 2.5\,K (Fig.~\ref{fig:ADR_crystal_growth}A). The Ne gas travels through the cryostat via a series of stainless steel and copper tubes that cool the gas from room temperature to slightly above the melting point ($\sim27$\,K). After the gas exits the tube at the final cooling stage, it begins condensing on top of the superconducting resonator. 
We use alkali atoms for atomic impurities as they have hyperfine transitions in the GHz regime; here, we work with sodium (Na). To create an atomic beam of Na, we heat an oven to $\sim170$\,$^{\circ}$C in a separate chamber directly attached to the outer vacuum shield. Na atoms reach the resonator through the holes in the 50\,K and the 4\,K shield. A shutter on the 4\,K shield helps to reduce the black-body radiation at the device when the crystal is not growing [see materials and methods in the supplementary materials; Fig.~S1]. 

We characterize the crystal deposition by continuously monitoring the resonator's frequency ($\omega_{\text{c}}$) during the growth time (Fig.~\ref{fig:ADR_crystal_growth}B). The resonance shifts down from its original frequency in the vacuum when the crystal growth begins ($t = 0$) due to the change in the effective dielectric constant ($\epsilon_{\text{Ne}} = 1.244$ \cite{cole1969image}) until the crystal completely covers the resonator's mode volume. Our niobium (Nb) resonator has a spiral design creating a homogeneous magnetic field that extends to approximately 120\,$\mu$m perpendicular to the surface \cite{Weichselbaumer2019spiral}. The resonator frequency shift is fitted as a function of the crystal thickness using an electromagnetic field solver software (Sonnet). Typical growth rates of the doped crystal are $\sim140$\,$\mathrm{nm/min}$ [see materials and methods in the supplementary materials; Fig.~S2 and Fig.~S3].

To verify that Na atoms are embedded in the crystal, we perform absorption spectroscopy using two split fibers, which are positioned approximately 1\,cm above the superconducting chip at a 45$^{\circ}$ angle. The input fiber couples broadband light into the cryosystem, while the output fiber is connected to an optical spectrometer (OS) (Fig.~\ref{fig:ADR_crystal_growth}A). The optical transmission, T, that we observe is described in terms of the optical depth (OD), which is defined as $\mathrm{T \equiv e^{-\text{OD}}}$. We normalize the OD by subtracting a linear offset based on the OD at the off-resonant wavelengths 450\,nm and 800\,nm. The nonzero optical density that we observe at 100\,mK indicates the presence of trapped atomic Na (Fig.~\ref{fig:ADR_crystal_growth}C).

\begin{figure*}[t!]
    \centering
    \includegraphics[width=2\columnwidth]{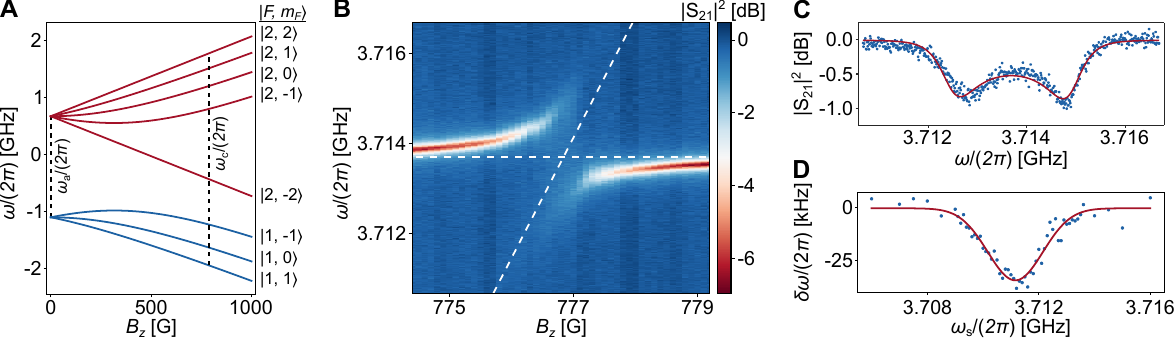}
    \caption{(\textbf{A}) Ground state hyperfine structure of Na $3^{2}S_{1/2}$ as a function of an external magnetic field. Black dashed lines indicate the ground state zero field hyperfine splitting of atomic Na ($\omega_{\text{a}}/(2\pi)$ = 1.771\,GHz) and the resonator frequency ($\omega_{\text{c}}/(2\pi)$ = 3.713\,GHz). We couple the $\ket{1,1}\rightarrow\ket{2,2}$ transition to the superconducting resonator. (\textbf{B}) Measured transmission spectrum of a resonator coupled to a Na ensemble, $|S_{21}|^{2}$, as a function of an external magnetic field, $B_{z}$, and probe frequency $\omega$ at 50\,mK. White dashed lines represent the cavity center frequency without atoms and the atomic transition with the modified hyperfine constant (-1.6\,$\%$) in an external magnetic field. (\textbf{C}) Transmission spectrum of a resonant Na ensemble ($\omega_{\text{a}}$ = $\omega_{\text{c}}$) as a function of probe frequency $\omega$ measured at a magnetic field of 776.95\,G and 50\,mK. (\textbf{D}) Spin distribution of atomic Na ensemble. Resonator frequency shift, $\delta\omega$, is measured as a function of spectroscopy tone frequency $\omega_{\text{s}}$ at 85\,mK and fitted with Gaussian distribution function.} 
    \label{fig:strong_coupling}
\end{figure*}

We attribute the dominant absorption line centered around 586\,nm to atomic Na's D1 and D2 line \cite{steck2024sodium} (red dashed line,  Fig.~\ref{fig:ADR_crystal_growth}C). Additionally, we see the expected spectral shifts, line broadening, and triplet splitting of the $S \rightarrow P$ transition \cite{Pipkin_ArRb, battard2023cesium}. These properties are attributed to the interaction between the Na atom and the crystal-field potential and have been observed for other alkali atoms and cryogenic gas matrices \cite{Xu2011YtNe, kanagin2013optical,upadhyay2016longitudinal,gaire2019subnanometer}. Furthermore, we observe the change in OD due to optical bleaching, which is also observed in other alkali-cryogenic gas matrices [see Fig.~S4 in the supplementary materials]. We estimate from the integrated OD that the density of atomic Na in the crystal is on the order of $\mathrm{3 \times10^{16}\,cm^{-3}}$. 

\section{Strong coupling of the spin ensemble}\label{sec2}

At $\sim3$\,K Na atoms are nearly equally distributed between the $F = 2$ and $F = 1$ hyperfine manifolds (Fig.~\ref{fig:strong_coupling}A). We reduce the temperature of the system to 50\,mK, where over 70\% of spins are polarized in the $F = 1$ hyperfine ground state. To further enhance this polarization and couple individual magnetic sublevels to the resonator, we use a pair of superconducting Helmholtz coils. 
The large magnetic field generated by coils splits the hyperfine manifold due to the Zeeman interaction, which is described by the Hamiltonian $H_{B} = -g\mu_{0}B_{z}S_{z}$.
We tune the atoms from their zero-field splitting ($\omega_{\text{a}}/(2\pi)=1.771$\,GHz, \cite{steck2024sodium}) to the resonator's frequency ($\omega_{\text{c}}/(2\pi)=3.714$\,GHz), which results in a 92\% polarization into the $F=1$ ground state. 
The magnetic sublevel with the largest spin polarization of approximately $34$\% is the $\ket{F,m_{f}}$ = $\ket{1,1}$ state. Our two-level system is defined by the transition between the $\ket{1,1}$ and the $\ket{2,2}$ hyperfine states. 

A single Na atom located 5\,$\mu$m above the resonator has the coupling strength of $g_0/(2\pi)$~$\approx 5$\,Hz, which is much smaller than the losses of the resonator ($\kappa/(2\pi)$ = 265\,kHz) and, hence, far from the strong coupling regime that would allow effective control of the Na spin with microwaves. We overcome this issue by coupling atoms collectively. The effective coupling is then enhanced, $g_{\text{eff}} = \sqrt{N}g_0$, with an increased number of atoms $N$ as predicted in the Tavis-Cummings model \cite{TavisCummings1968, fink2009dressed}. After optically bleaching the crystal completely [see materials and methods in the supplementary materials], we measure the transmission spectrum as a function of the applied magnetic field near resonance ($\omega_{\text{a}} \approx \omega_{\text{c}}$) and witness an avoided crossing at $B_z = 776.95$\,G (Fig.~\ref{fig:strong_coupling}B). We fit the vacuum Rabi splitting (Fig.~\ref{fig:strong_coupling}C) with the function derived from the steady-state Maxwell-Bloch equations and extract the effective coupling strength of $g_{\text{eff}}/(2\pi) = 0.95$\,MHz and the characteristic width of the inhomogeneous atomic ensemble of $\Gamma/(2\pi) = 716$\,kHz [see materials and methods in the supplementary materials].

The extracted atomic linewidth is in good agreement with the one obtained in the direct measurement of the spin distribution for a similar Ne crystal doped with Na (Fig.~\ref{fig:strong_coupling}D). There, we tuned the Na atoms $\sim10$\,MHz away from the resonator and applied narrow ($\Gamma_{\text{pulse}} \ll \Gamma)$ microwave pulses to the spin ensemble while simultaneously monitoring the resonator's frequency. As the microwave pulses scan across the spin sub-ensemble, we witness a shift in the resonator's frequency proportional to the number of Na atoms saturated. The obtained spin distribution is fitted with a Gaussian distribution function, and the extracted atomic linewidth is $\Gamma/(2\pi) = 989$\,kHz. We note that the simulated atomic linewidths due to dipole-dipole interactions from Na-Na result in a linewidth $\sim 300$\,kHz; similarly, the dipole-dipole interaction from Na-$^{21}\text{Ne}$ is $\sim 0.36$\,kHz. We attribute the additional broadening to the likely combination of the inhomogeneous interaction between the atoms and the host matrix, and dipole-dipole coupling to unwanted paramagnetic impurities such as $\mathrm{Na_{3}}$.  

The cooperativity of our system, $\mathrm{C} = g^{2} / (\kappa \Gamma)\approx 9$,  demonstrates that we operate in the strong-coupling regime \cite{kimble1998strong}, enabling high-fidelity coherent exchange between the atoms and the superconducting resonator, which is essential for quantum sensing and quantum information processing applications.

\begin{figure*}[t]
    \centering
    \includegraphics[width=1.9\columnwidth]{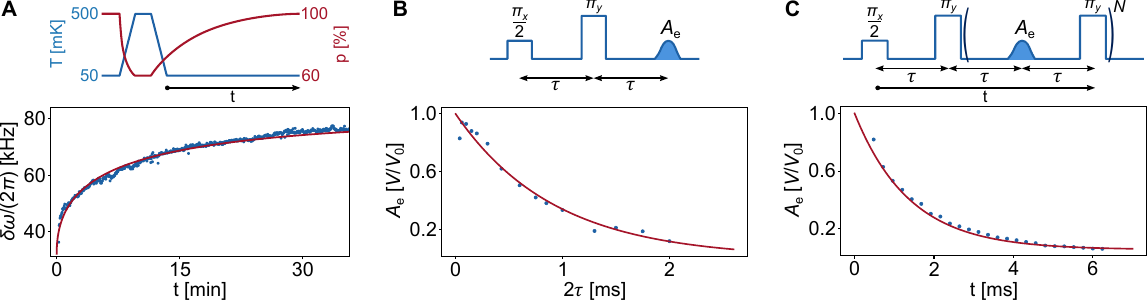}
    \caption{(\textbf{A}) Longitudinal relaxation of spins. Resonator shift, $\delta\omega$, as a function of time after ramping down the temperature of the spins and the cavity. The decay time is fitted with a stretched exponential function, $T_{1}$ = 8.2\,min. Inset describes the temperature of the spin ensemble and cavity (blue) and population of the spins in the hyperfine ground state $\ket{F}$ = $\ket{1}$ during the measurement. (\textbf{B}) Normalized echo amplitude $A_{\text{e}}$ as a function of delay $\tau$ between the $\pi/2$- and the $\pi$-pulses. An exponential function fit yields a coherence time $T_{2,\text{Hahn}}=922$\,$\mu$s. The inset shows the Hahn echo sequence. (\textbf{C}) Normalized CPMG echo amplitude $A_{\text{e}}$ as a function of the measurement time after the $\pi/2$ pulse. Each data point is averaged over 20 measurements. An exponential function fit results in a coherence time $T_{2,\text{CPMG}}=1.38$\,ms. The inset shows the CPMG sequence with $N = 25$ $\pi$-pulses.   
}
    \label{fig:coherencetimes}
\end{figure*}

\section{Longitudinal relaxation and coherence measurements}

We measure spin ensemble relaxation and coherence times to benchmark our hybrid platform for future applications. We start by measuring the longitudinal relaxation time $T_1$ as a characteristic time needed for a non-equilibrium state to decay to its thermal equilibrium value \cite{astner2018solid}. We begin by detuning Na atoms $\Delta/(2\pi)=(\omega_{\text{c}}-\omega_{\text{a}})/(2\pi)=28$\,MHz $\gg g_{\text{eff}}/(2\pi)$ from resonance to suppress the Purcell enhancement of the single-spin spontaneous emission rate \cite{purcell1995spontaneous, bienfait2016controlling}. Then, we increase the temperature of our system to 500\,mK, where spins are in a roughly equal mixture of $F=1$ and $F=2$ states, 58\% in $F=1$.  After the system reaches a thermal steady state, we non-adiabatically switch the temperature back to 50\,mK at a rate of 100\,mK per minute to create a non-equilibrium state of spins and monitor the resonator's frequency shift $\delta\omega$ as the spin system re-polarizes (Fig.~\ref{fig:coherencetimes}A). The measured decay profile of the dense ensemble is fitted with a stretched exponential function ($e^{-\sqrt{t/T_1}}$) \cite{choi2017depolarization}, and results in a relaxation time $T_{1}= 8.23$\,min. 

Next, we measure the coherence time $T_{2}$ of the atomic ensemble, starting with a Hahn echo pulse sequence ($\pi_x/2 - \tau - \pi_y - \tau  - A_{e}$). We use a weak 1\,$\mu$s rectangular $\pi/2$-pulse to create a transverse magnetization in the spin ensemble, followed by a strong refocusing $\pi$-pulse of the same duration after the delay time $\tau$. The echo ($A_{e}$) appears at time 2$\tau$ after the initial $\pi/2$-pulse (inset, Fig.~\ref{fig:coherencetimes}B). The normalized amplitude of the echo pulse plotted as a function of the delay time is fitted with an exponential decay $e^{-2\tau/T_{2,\text{Hahn}}}$, where $T_{\text{2,Hahn}}=0.92$\,ms (Fig.~\ref{fig:coherencetimes}B). 

We can extend the spin coherence time by employing dynamic decoupling with the Carr-Purcell-Meiboom-Gill (CPMG) pulse sequence \cite{CPMG1954, Meiboom1958ModifiedSM}. In addition to the $\pi/2$- and $\pi$-pulses in the Hahn echo scheme, we include $N=25$ $\pi$-pulses spaced 2$\tau$ from the original $\pi$ pulse (inset, Fig.~\ref{fig:coherencetimes}C). The delay time $\tau$ is fixed to 100\,$\mu$s. CPMG echo sequence is robust against pulse imperfections, and the echo decay time can be measured in a single sequence. We repeat the measurement 20 times and plot the normalized average echo amplitude as a function of time, fit it to an exponential decay, and obtain a coherence time $T_{2,\text{CPMG}}= 1.38$\,ms, which is 50\,\% longer than $T_{2,\text{Hahn}}$ (Fig.~\ref{fig:coherencetimes}C). 

\section{Discussion and outlook}\label{sec12}

We have achieved an effective collective coupling rate $g_{\text{eff}}/(2\pi) = 1.19$\,MHz and coherence times $T_{2,\text{CPMG}}= 1.38$\,ms, demonstrating the viability of this platform for coherent spin control. Our study shows that impurities embedded in cryogenic solids atop superconducting resonators serve as an exciting platform for quantum information sciences and future quantum optics experiments. 

There remain several clear pathways for improving system performance. Our current initialization of the system in the single hyperfine state ($\ket{F,m_{f}}$ = $\ket{1,1}$) relies solely on thermal polarization at 50\,mK, where the thermal population of this state is around 34\,\% at a large magnetic field. Higher population in a single state can be achieved by going to lower temperatures, implementing optical pumping techniques \cite{kanagin2013optical,lang1995hyperfine}, or selecting impurities with a reduced number of $m_{F}$ states similar to hydrogen. A more efficient initialization scheme will also boost the effective coupling by increasing the number of participating impurities. 

Larger coupling can also be obtained using a microwave resonator with higher mode volume, increasing atomic density, or choosing an impurity with a larger magnetic dipole moment, such as dysprosium ($ \sim 10\mu_{B}$). Moving beyond magnetic coupling, electrically coupled impurities, such as polar molecules \cite{PolarMoleculesDeMille}, could reach single-dopant strong coupling at megahertz rates when integrated with appropriately designed resonators, enabling single impurity operation with microwaves. 

The coherence time of the system is limited primarily due to the presence of non-zero spin $^{21}$Ne nuclei in the crystal lattice ($T_{\text{2,Na-$^{21}$Ne}}\approx2.8$\,ms). Longer and more sophisticated dynamical decoupling methods could further extend the coherence of the atoms \cite{Ajoy2011UhrigDecoupling, bar2013solid}. Additionally, this loss mechanism can be fully suppressed by using isotopically purified neon gas or by changing the host matrix material.

Overall, our results establish doped cryogenic solids integrated with superconducting circuits as a flexible, high-coherence platform for hybrid quantum technologies, with broad potential applications in quantum information processing, quantum sensing, and engineered quantum optics.

\begin{acknowledgments}
We would like to thank Igor Mazets for fruitful discussions and Aslihan Sasmaz for optimizing crystal deposition parameters for the follow-up measurements. This work is supported by the European Research Council: ERC-AdG Emergence in Quantum Physics (EmQ) under Grant Agreement No. 101097858. Furthermore, we acknowledge support by the Austrian Science Fund (FWF): COE 1 Quantum Science Austria and project P34314 (Spins in Quantum Solids). E.S.R. has received funding from the FWF and Austrian Academy of Science (\"{O}AW) project DI\_2023-007\_REDCHENKO\_HQCP. This work forms the basis of a patent application (PCT/EP2025/066647; filed June 13th, 2025).
\end{acknowledgments}


%

\onecolumngrid
\newpage
\normalsize
\appendix
\begin{center}
\textbf{\large{Supplementary Materials for: "Impurities in cryogenic solids: a new platform for hybrid quantum systems"}}
\end{center}

\section*{Materials and Methods}

\subsection*{Crystal creation setup}

The crystal creation experimental setup is based on the design from Ref.~\cite{kanagin2018design} using an adiabatic demagnetization refrigerator (ADR) purchased from STAR Cryoelectronics~\cite{StarCryo}. The crystal growth begins with opening the host gas tank of natural neon (Ne), which is a combination of $90.48\%$~$^{20}$Ne (spin-0), $9.25\%$~$^{22}$Ne (spin-0), and $0.27\%$~$^{21}$Ne (spin-3/2), and regulating the pressure with a needle valve as the gas enters the cryosystem [Fig.~\ref{adr_test}]. Ne travels through a thin stainless steel tube, which is connected to a copper tube heat anchored to the first stage of the pulse tube, where it is cooled from 300\,K to 50\,K. As the gas exits the copper tube, it again goes through a stainless steel tube that acts as a thermal break between the first and the second stages. Ne then enters the second copper tube, which is weakly connected to the second stage via a heat link and is regulated at a temperature slightly above the Ne melting point (27\,K). The final stainless steel tube connected to the second copper tube is aimed directly at the superconducting resonator mounted at the mK stage. The ADR allows us to precisely control the temperature of the mK stage during the crystal growth and analysis during the regulation cycle. The ADR has the capability to grow thin ($\mathrm{\approx 1}$\,$\mu$m) crystals at mK temperatures and thick ($\mathrm{\gg 1}$\,$\mu$m) crystals below the base temperature of the pulse tube. Our Ne crystals are typically grown at temperatures ranging from  2.3\,K to 2.5\,K for 2 hours. Once we have used the entire magnetocaloric cooling power of the ADR, we thermally anchor the mK stage with the second stage of the crysostat that is typically at a temperature of 2.9\,K. The low deposition temperatures we use result in higher optical quality crystals, leading to increased optical polarizations \cite{kanagin2013optical, pathak2014, upadhyay2019spin}. Additionally, the mK stage remains below 3.3\,K when cooling to 50\,mK, which minimizes thermal annealing and melting of the host crystal.  

In order to dope the Ne crystals, we heat up an atomic oven directly attached to the 300\,K shield. The oven is filled with a 5\,g capsule of atomic sodium (99.95\%). The body of the oven is heated above the melting point of 371\,K (97\,$^\circ$C) and regulated to a temperature of 433\,K (160\,$^\circ$C); we keep the front of the oven hotter by 5\,K to reduce condensation of the alkali vapor at the 1\,mm nozzle, where the atoms exit from. During the warm-up, the oven is located approximately 20\,cm away from the 50\,K shield. Once Ne begins to condense, we push the oven forward with a linear actuator to the 50\,K shield, such that the nozzle is roughly 3\,mm away from the shield. We then open the shutter on the 3\,K stage and begin depositing on the resonator simultaneously with the noble gas. We monitor the deposition process by periodically measuring the superconducting resonator.

Once we have completed the crystal growth, we stop the temperature regulation of the second copper tube and the oven. We close the shutter and pull the oven back into the atomic chamber. The copper tube which was held at 27\,K cools to 6\,K. To insure the presence of impurities, we characterize the crystal optically at $\sim3$\,K using absorption spectroscopy. 

\subsection*{Superconducting resonator}

\begin{figure}[t]
    \centering
    \includegraphics[width=.77\columnwidth]{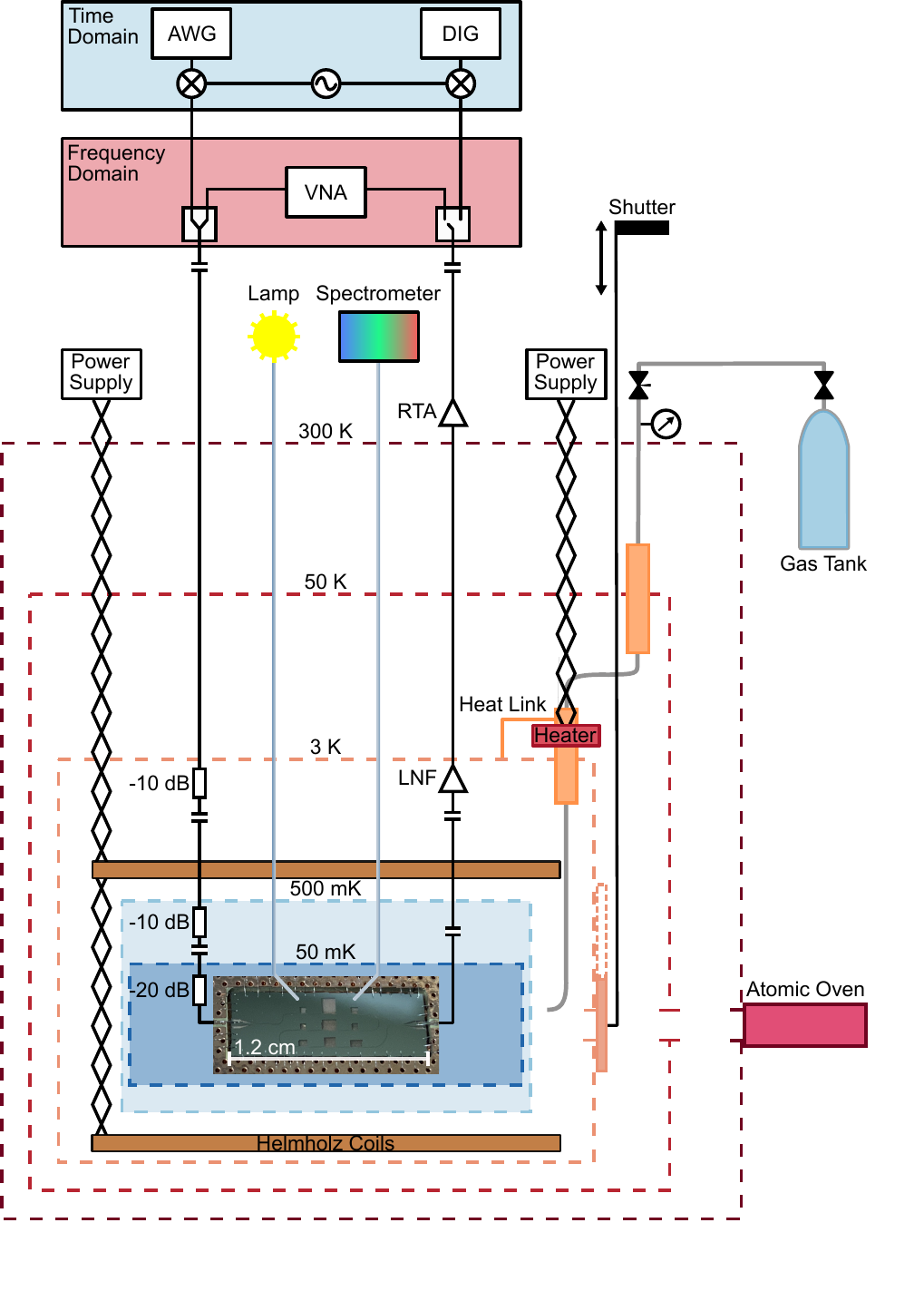}
    \caption{Schematic of the cryogenic setup for the crystal growth including optical and microwave spectroscopy setups. The thermalized shields are positioned at 300\,K, 50\,K, and 3\,K stages. The holes in the 300\,K, 50\,K, and 3\,K shields are aligned with the atomic impurities oven and the resonator position.   }
    \label{adr_test}
\end{figure}

\begin{figure}[t]
    \centering
    \includegraphics[width = 0.95\columnwidth]{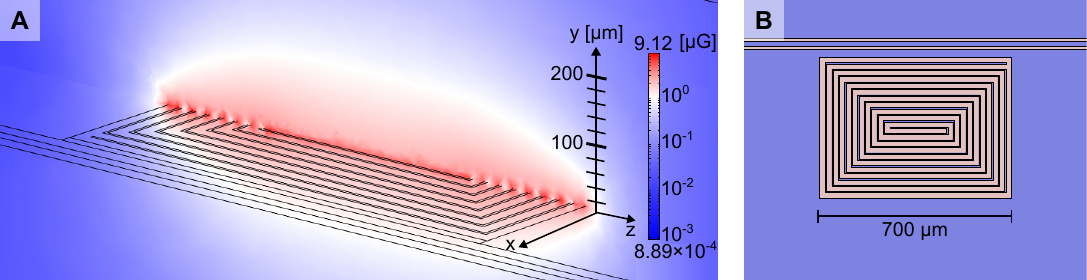}
    \caption{(\textbf{A}) COMSOL simulation of the magnetic field extending out in the direction of the y-axis, where the z-axis is defined by the field produced by the superconducting Helmholtz coils. (\textbf{B}) Schematic of the spiral resonator.  }
    \label{resonator}
\end{figure}

The superconducting resonator is patterned on 330\,$\mathrm{\mu m}$-thick sapphire wafer using 200\,nm niobium (Nb) layer. In addition to the $\sim3.7$\,GHz resonator, we have 8 spiral resonators at frequencies ranging from 1.7\,GHz to 11\,GHz that can be seen on the chip in Fig.~\ref{adr_test}. The magnetic field of the spiral resonator extends approximately 120\,$\mu$m out of the plane [Fig.~\ref{resonator}A], offering relatively homogeneous coupling within the resonator's mode volume \cite{Weichselbaumer2019spiral}. 

\begin{figure}[t]
    \centering
    \includegraphics[width = 0.55\columnwidth]{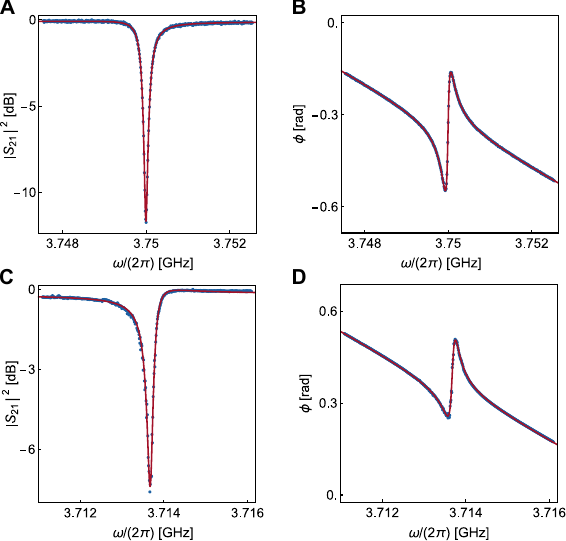}
    \caption{Transmission spectrum, $|S_{21}|^2$, as a function of frequency, $\omega/(2\pi)$ of the resonator before (\textbf{A}) and after (\textbf{C}) the crystal growth. Phase response, $\phi =\arctan(\Im(S_{21}),\Re(S_{21}))$, as a function of frequency, $\omega/(2\pi)$ of the resonator before (\textbf{B}) and after (\textbf{D}) the crystal growth. Blue dots show measured data, red line is the fitted function Eq.~\ref{eq:S21_IO}.  }
    \label{resonator_meas}
\end{figure}

In addition to coupling to atoms, the resonator is used to measure the thickness of the crystal while growing. The crystal deposition temperatures ranging from 2.3 to 2.9\,K are well below the critical temperature of Nb ($\mathrm{T_{c}}$ = 9.3\,K). As the gas condenses on top of the resonator, the fundamental frequency shifts due to a change in the participation ratio between the sapphire substrate, Nb, vacuum, and solid Ne. We start with the bare resonator at $\omega_{\text{c}}/(2\pi)=3.75$\,GHz, $\kappa_{\text{e}}/(2\pi) = 195$\,kHz, $\kappa_{\text{i}}/(2\pi) = 70$\,kHz, once the resonator mode volume is filled the resonator shifts down to $\omega_{\text{c}}/(2\pi)=3.7137$\,GHz, $\kappa_{\text{e}}/(2\pi) = 163$\,kHz, $\kappa_{\text{i}}/(2\pi) = 118$\,kHz, see Fig.~\ref{resonator_meas}. The frequency-dependent transmission $S_{21}$ has been fitted using the following function \cite{probst2015efficient}:

\begin{equation}\label{eq:S21_IO}
    S_{21}(\omega) = A e^{i(\alpha - \omega \tau)} \left(1-\frac{\kappa_{\text{e}}e^{i \psi}}{\kappa_{\text{e}}+\kappa_{\text{i}}+2i(\omega - \omega_{\text{c}})}\right),
\end{equation}
where A is the background transmission amplitude, $\alpha$ is the phase offset, $\tau$ is the microwave delay, $\omega$ is the signal frequency, $\omega_{\text{c}}$ is the resonance frequency, and $\psi$ is the phase shift of resonance due to impedance
mismatch with the transmission line. 
The increase in the intrinsic resonator losses is attributed to the presence of atomic dopants in the crystal. The frequency shift as a function of Ne thickness is simulated using the Sonnet software. We model Nb as a lossless metal, and add a uniform dielectric layer ($\epsilon_{\text{Ne}}$ = 1.244) of Ne directly on top with a variable thickness [see Fig.~1 in the main text].

\subsection*{Theoretical description of the system}

We measure the inhomogeneously broadened spin distribution in the dispersive regime, Fig.~2D, for a similar density crystal. This distribution is well described by a Gaussian function:
\begin{equation}\label{eq:pho}
\rho(\omega)=b-a e^{\frac{-(\omega-\omega_\mathrm{a})^2}{2\sigma^2}},
\end{equation}
where $a$ and $b$ are normalization parameters, $\omega_\mathrm{a}$ is the spin center frequency, $\gamma_q=2\sqrt{2\mathrm{ln}2}\sigma$ defines the FWHM of the distribution.

To theoretically model an inhomogeneously broadened atomic ensemble, we discretize the spin frequency distribution $\rho(\omega)$ into $N_\rho \approx 2000$ spin packets with equal spacing. Each of the $N_\rho$ numerical spins, indexed by $j$, couples to the cavity with coupling strength $g_\rho^j = g_\mathrm{coll} \rho_j$ and is weighted according to $\sum_j\rho_j=1$. The coupled system is then described by the Tavis-Cummings Hamiltonian \cite{tavis1968exact},
\begin{align}
\begin{split}
    \mathcal{H} = &\hbar \omega_\mathrm{c} a^\dagger a + \frac{\hbar}{2} \sum_i \omega_\mathrm{s}^j \sigma_z^j 
    + i \hbar \sum_j g_\rho^j \left(a^\dagger \sigma_-^j - \sigma_+^j a \right),
\end{split}
\end{align}
where $\omega_\mathrm{c}$ is the cavity frequency, and $\omega_\mathrm{s}^j = \omega_\mathrm{a} + \Delta_s^j$ are the frequencies of the numerical spin packets. 

The system's losses caused by coupling to the environment are accounted for by introducing a cavity loss rate $\kappa = \kappa_{\text{i}}+\kappa_{\text{e}}$ and a spin decoherence rate $\gamma_\perp$. The transmission spectra for vacuum Rabi splitting [Fig.~2C, Fig.~\ref{OD_comparison}D] are fitted using the function derived from the steady-state Maxwell-Bloch equations with the Fano for the notch-type resonator (Eq.~\ref{eq:S21_IO}):
\begin{equation}\label{eq:S_21_Nat}
S_{21}(\omega) = A e^{i(\alpha - \omega \tau)}\left(1-\frac{\kappa_{\text{e}}/2e^{i \psi}}{\kappa/2+i(\omega - \omega_{\text{c}})+\sum_{j=1}^{N_{\rho}}\frac{{g_\rho^j}^2}{\gamma_\perp+i(\omega-\omega_\mathrm{s}^j)}}\right).
\end{equation}
We normalize the data for the Fig.~2C in the main text and fix all the resonator parameters ($\kappa$, $\kappa_\mathrm{e}$, $\omega_\mathrm{c}$) from the separate high-power measurement, leaving $\gamma_\perp$, $\gamma_q$, $\omega_\mathrm{a}$, and $g_\mathrm{coll}$ as free parameters.

The effective linewidth $\Gamma=1/T_2^*$, from the definition of the cooperativity $\mathrm{C} = g^{2} / (\kappa \Gamma)$ in the main text, can
be calculated using \cite{julsgaard2012dynamical}:
\begin{equation}
    \Gamma = \left[\int_{-\infty}^{{-\infty}}\frac{\rho(\omega)d\omega}{\gamma_\perp+i(\omega-\omega_\mathrm{a})}\right]^{-1} \approx \left[ \sum_{j=1}^{N_{\rho}} \frac{\rho_j} {\gamma_\perp + \frac{{\Delta_\mathrm{s}^j}^2}{\gamma_\perp} }\right]^{-1}.
\end{equation}

\subsection*{Optical bleaching at mK temperatures}

\begin{figure}[t]
    \centering
    \includegraphics[width = 0.8\columnwidth]{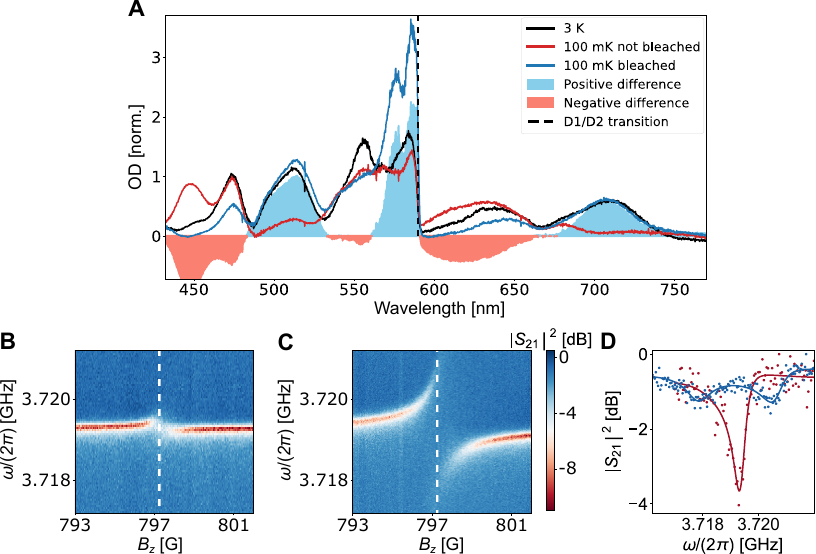}
    \caption{(\textbf{A}) Absorption spectroscopy of the crystal at different temperatures: normalized OD as a function of wavelength at the substrate temperature of 3\,K (black) and before/after optical bleaching at 100\,mK (red/blue). The filled-in areas demonstrate the positive (light blue) and negative (light red) differences in OD. (\textbf{B, C}) Measured transmission spectrum of a resonator coupled to an unbleached/bleached Na ensemble $|S_{21}|^{2}$ as a function of an external magnetic field $B_z$ and probe frequency $\omega$ at 50\,mK. White dashed line identifies the position of vacuum Rabi splitting. (\textbf{D}) Vacuum Rabi splitting measured in transmission spectra $|S_{21}|^{2}$ for the resonant ($\omega_{\text{c}}=\omega_{\text{a}}$) bleached (blue) and unbleached (red) Na ensemble, $|S_{21}|^{2}$, as a function of an external magnetic field and probe frequency $\omega$ at 50\,mK.  }
    \label{OD_comparison}
\end{figure}

We observe an optical bleaching, or optical annealing, effect that is common for impurities in cryogenic crystals \cite{kanagin2013optical} [Fig.~\ref{OD_comparison}A]. For bleaching, we use white light, which is generated by two LEDs (447\,nm and 517\,nm peak wavelength) and a tungsten lamp, coupled into an optical fiber that enters the cryostat. The end of the fiber is attached to the sample box at a 45$^\circ$ angle to the substrate. Enhanced bleaching is observed in our system at temperatures below 2\,K, where the power emitted from the end of the fiber is $\sim 90$\,$\mu$W and illuminates the crystal for roughly 15 seconds. The average normalized optical depth (OD) of the atomic Na (550-588\,nm) reaches roughly 1.4 at 3\,K [Fig.~\ref{OD_comparison}A, black line]. Reducing the substrate temperature from 3\,K to 50\,mK alone does not drastically change the optical spectrum of the doped crystal [Fig.~\ref{OD_comparison}A, red line]. Only after we apply white light a dramatic change in the OD occurs and we observe an increase in OD of approximately 1.5 at a wavelength around 586\,nm [Fig.~\ref{OD_comparison}A, blue line]. We noticed no significant changes in the OD after bleaching at 50\,mK for up to 12 hours, which is the limit of our ADR capabilities. Once the substrate temperature reaches 3\,K, the system returns to the original  3\,K spectrum. 

Bleaching at mK temperatures has a drastic effect on the coupling between the spin ensemble and the cavity [Fig.~\ref{OD_comparison}B,C]. The coupling strength of an unbleached ensemble is only $g_{\text{UB}}/(2\pi) = 428$\,kHz [Fig.~\ref{OD_comparison}D], which is almost 3 times lower than the coupling we achieved for the fully bleached Na ensemble ($g_{\text{eff}}/(2\pi) = 1.19$\,MHz). Hence, there is a correlation between the number of Na atoms observed in the absorption and microwave spectroscopies. 

We believe the increased OD between the 3\,K and 50\,mK is due to an increased absorption into the zero-phonon line. Similar effects have been witnessed in other solid-state systems with vacancies and impurities \cite{Tinvacancy_iwasaki, ZPL_diamond_Zhang}.

\newpage
\clearpage
\newpage

\clearpage
\newpage

\clearpage
\newpage

\clearpage  

\end{document}